# Integrating users' needs into multimedia information retrieval system

MAGHREBI Hanène (PhD student, DAVID Amos ( Professor)

LORIA- Campus Scientifique, B. P. 239, 54506, Vandoeuvre-lès-Nancy, France

Hanene.Maghrebi ; Amos.David} @loria.fr & @loria.fr

**Abstract:**

The exponential growth of multimedia information and the development of various communication media generated new problems at various levels including the rate of flow of information, problems of storage and management. The difficulty which arises is no longer the existence of information but rather the access to this information.

When designing multimedia information retrieval system, it is appropriate to bear in mind the potential users and their information needs. We assumed that multimedia information representation which takes into account explicitly the users' needs and the cases of use could contribute to the adaptation potentials of the system for the end-users. We believe also that responses of multimedia information system would be more relevant to the users' needs if the types of results to be used from the system were identified before the design and development of the system.

We propose the integration of the users' information needs. More precisely integrating usage contexts of resulting information in an information system (during creation and feedback) should enhance more pertinent users' need. The first section of this study is dedicated to traditional multimedia information systems and specifically the approaches of representing multimedia information. Taking into account the dynamism of users, these approaches do not permit the explicit integration of the users' information needs. In this paper, we will present our proposals based on economic intelligence approach. This approach emphasizes the importance of starting any process of information retrieval witch the user information need.

**Introduction**

The aim of a multimedia information retrieval system is to provide answers that correspond best to the users information needs. Generally, in information systems (IS), users express their information needs in form of queries and then the system matches the formulated queries to the database to find relevant information. In the case of multimedia information, the information needs of users can be related to the document in its totality or to its sub-sets.

For multimedia information, the information needs can correspond to multi-media information in its entirety or to an audio-visual component, for example the sequences, the scenes, the passages, the dialogs… In multimedia information bases we try to facilitate the access to the relevant bits of information which the user needs and to adapt the answers of the system to diversified information needs. The questions, which can be asked and to which we try to provide answers through economic intelligence approach, are the following:
- How do we facilitate users' access to multimedia information?
- How can the informational needs of users be better met?

We try to elaborate on an adaptative information system that is able to provide relevant information to users informational needs according to their context of use.

In our approach, we propose to represent users in the information system during the design phase. Users are represented by integrating their information needs into "context of use" of information search results. Our information system also put into consideration the different level of granularity of multimedia documents.

**1.     Multimedia information retrieval systems (MIRS)**

Information retrieval concerns the representation, storage, organisation and access to information. A multimedia information system is an IS that takes into account the information type, its characteristics and components (image, sound, text) in order to allow users to have access to these information.

According to David (2005), 'Information Systems, in general, have experienced some profound evolutions either in the objective of use – exploitation – or in the nature (typology) of managed information….' The nature or the type of information that the IS manages, influences the functionalities of the system.

Information retrieval is a process or method by which a user is able to convert its information needs into a list of queries that should furnish (as results) documents containing information that respond to its needs.  Information retrieval includes the intellectual aspects of information description and specification of the totality of the system. Moores (1951)

Research works on information retrieval are carried out using various approaches. Classical system oriented approach were being used until early 80s. This was followed by cognitive user-oriented approach in the 80s. Since then, the interest is no more on the system but on users and their interactions with the systems.

Indeed, during the information retrieval process, the user interacts with the IS. This interaction (human-machine) makes one interested in the data at the detriment of the user. As a result, user and document modelling techniques were introduced into the studies of information retrieval system. These techniques follow the document typology and the nature of users' information requirements Saracevic (1995, 1996). The Human-Machine Interaction (HMI) takes a new form and then becomes Human-Machine-Human interaction (HMH) David (2006). David explained 'that all the retrieval methods suggested in IRS deprived of user modelling rest on the capacity of the user to express his needs and in particular to organise his knowledge into hierarchy' David (1999). However it is not obvious for a user to describe his information needs or to translate them into queries. This leads to the third approach which is use-oriented in which researchers' attention are put on the study of the user and his environment.

The current research on this last approach 'users studies' considers that the user of information engages in an information search behavior in answer to a perceived need. The need concept seems to be very difficult to translate into research queries whereas the information-seeking behavior concept is an activity (information related activity) identifiable, observable and being able to be deconstructed.

Wilson (2000) assumed that it is sometimes unclear witch of the senses is meant the researcher had in mind when setting the research objectives. The professional perceives a need, but he cannot always define it or know who is in need.

In information retrieval systems and more precisely in information processing, information representation is paramount.

According to Eisenberg and Berkowitz (1992), the information retrieval problem relates to the organization, the presentation of information and also the definition of the task, the identification of information needs, the putting in place of information retrieval strategy, and the use of information. Belkin (1993, 1995), in his approach supposed that the real problem of information retrieval process is not the representation of informational objects but rather the representation of the Anomalous State of Knowledge (ASK) of the user.

Following this presentation, we assumed that integrating the information needs of user in information system, will contribute to the improvement of the interaction of the user with the system and there will be more possibility in meeting the information needs of the user.

## 2. Integrating the information needs of user in information system in order to make it adaptive

To our knowledge, multi-media information systems do not explicitly represent information needs of user. It is rather the representation of the user in the form of profile which is presented. The starting point of our approach is the users' information need. We integrate the informational needs of user in information system through his representation. We explicitly take into account the user and his needs just as the representation of the characteristics of document and its contents.

### 2.1 Users' information needs

In information retrieval field, the concept of information need is of great importance because information retrieval is undertaken with the aim of meeting some of these information needs (Dervin1986, Belkin 1980, Codiac 1998, Polity 2000, Knitting 2003).

The concepts related to information retrieval, for example the concept of relevance, also relate to the notion of information need. According to Tricot, 'information need is a necessity to fill up a noted deficiency in information, a gap, a defect or an anomaly…' The user expresses his information needs by queries. This stage in the operation of information system constitutes the first difficulty for the user.

According to Chevallet and Nigy (1996), the precise expression of information needs by a user is a form of knowledge of his research domain. It is not obvious for the user to translate his information needs into the system language.

In fact, the capacity of the user to clarify his information requirements depends on the level of his knowledge of the domain in which he makes his research and also on his knowledge of the information system Afolabi (2007). Various schools of thoughts have emerged in order to meet the information needs of users: Approach of 'need by negotiation' by Kuhulthau (1993 and 1999), approach of 'information practice' of Wilson (1999 and 2000).

Kuhulthau modelled the information retrieval process in six stages. This modelling highlights the dynamic nature of information requirements and the psychological states which accompany them. The six stages are: problem recognition, identity and formulation of this problem, data collection, presentation and evaluation of information.

Wilson studied the question of the access and the information retrieval under practical aspect (situation) information. From the Wilson approach and his nested model Wilson (1999), we can distinguish four practices, which we present in the following paragraphs.

Information Behavior is the totality of human behaviour in relation to sources and channels of information, including both active and passive information seeking, and information use. Thus, it includes face to face communication with others, as well as the passive reception of information as in? For example, watching TV advertisements, without any intention to act on the information given.

Information Seeking Behavior is the purposive seeking for information as a consequence of a need to satisfy some goal. In the course of seeking, the individual may interact with manual information systems (such as a newspaper or a library), or with computer – based systems (such as the Word Wide Web).

Information Searching Behavior is the 'micro–level' of behavior employed by the searcher in interacting with information system, whether at the level of human computer interaction (for example use of the mouse and clicks on links) or at the intellectual level ( for example, adopting a Boolean search strategy or determining the criteria for deciding which of two books selected from adjacent places on the library shelf is most useful ) which will also involve acts, such as judging the relevance of data or information retrieved.

Information Use Behavior consists of the physical and mental acts incorporating the information found in the person's existing knowledge base. It may involve, for example, comparison of new information with existing knowledge.

Wilson regards the practices of access to information (information seeking and use behaviour) as the set of actions and choices of actors put in place for information retrieval and use. Wilson's works are attached to the cognitive user-oriented paradigm era. This approach 'considers that attention should be put on the real needs of user and his environment'.

David (2003) approached the practical aspect of information retrieval by a knowledge acquisition model that consists of four phases: (a) exploration of the information world, (b) interrogation of the base of information, (c) analyzes of information base, (d) annotation based on the preferences and the individual discoveries.

American library association advanced a definition, which is interesting to us, of the informational need which puts forward the bond of the requirement in information and the use of information. 'To be competent in the use of information means that one can recognize when a need for information emerges and that one is able to find information adequate as well as evaluating and exploiting it'.

In Economic Intelligence context, the information needs would be a lack of information in a situation relative to a decisional problem. This decisional situation makes the user to begin a search for information and determines the process of information retrieval. We assume that the information needs of the user presuppose an expectation and a 'use' of retrieved information Maghrebi and David (2006).

*2.2    User modelling*

In our study, our reflection is based on the principle that the user seeks information for a particular use. We modelled the user by the contextualisation of the uses of information retrieval results. Thus, we defined a user model which contains the following information:
- User Identity (name, first name, address….)
- Information needs of user: they correspond to the contexts of use of retrieved information.

The information needs for the user are translated into contexts of use of retrieved information. The contexts of use of retrieved information could better reflect the real information needs of the user.

The integration of the information needs in the system is ensured by the representation of the contexts of use. In order to be able to satisfy the user's information needs, we represent the user and his information needs (in addition to the representation of the documents) in the database, in form of user profile and contexts of use of the retrieved information (Figure1).

The information needs are integrated in the system via attributes relating to the user and the contexts of use. These attributes can be predefined or added by the user at the time of search for information by annotation process. In the application we are developing to illustrate our proposals, we defined four contexts of use of retrieved information: training, teaching, entertainment, documentation Maghrebi (2007). These predefined cases of use are integrated in the information system during the development of the system. We are conscious that we might not be able to anticipate all the possible cases of use of retrieved information hence; it is possible that the user might not be able to find the attributes of the context relating to him. In order to make up for this problem of 'missing' contexts of use, we choose an open structure of representation of multimedia information. The developer of the system can integrate the new contexts which he considers important, at the time of system update.

Indeed, the user might not find among the attributes, that the system offers, the attribute which clarifies his information requirements or the attributes that the user esteem necessary to solve his information retrieval problem. To take this dynamic aspect of the contexts of use into account, we propose an approach allowing us to represent new contexts of use even after the development of the information system. This approach consists in making it possible for the users to specify new types of use through annotation process. This specification is expressed by the attributes regrouped on contexts of use. The user can also give interpretations to results obtained. The contexts of use of retrieved information can then be used by various users for the same informational needs or different needs. The use of the contexts of use can be regarded as a form of collaborative information retrieval. We refer here to Kurzke (1998) which regards the use of profiles as a form of collaborative research.

Case scenario 1: within the framework of a research team whose principal axis is economic intelligence, the video of the person in charge of economic intelligence (EI) at the national level can constitute the object of various information needs and various uses. Our document is made up of image, sound, text. Our representation of this document takes into account these components and differentiations of the user's needs in a documentation context, a member of the research team might have recourse to this document to prepare for collaboration with the person in charge of EI in the report. This first user needs to have an idea of how the subject was treated by the person in charge. The user will be able to go through the document and discover the evoked topics. This same audio-visual document can be an object of another use.

Case scenario 2: In a training context, another user (member of the same research team) seeks to argue an idea that he wants to raise in a scientific article. The user in this context needs precise quotations concerning French companies and the practice of the EI. The user can either consult the tree structure of the video, or choose sub-attributes of the training use context, and then choose the concerned level (scene, plan…).

In both cases, the object of information needs is the same (the audio-visual document) whereas the contexts of use of the document are not the same. The knowledge on contexts use can make document representation relevant compared to the user.
We represent the base of document and the user in this form: figure 2

**Conclusion:**

Our research proposes to integrate the informational needs of users into information system.
The multi-media information system on which we work could be regarded as an adaptive system since:
- It is user-use oriented
- It should make it possible to adapt or adjust its responses to the context of use of the user.

The process of annotation ensures an interaction renewed between the user and the system. Our system ensures this interaction by the integration of the informational needs of user on one hand and on the other hand the process of annotation which makes it possible for user to annotate retrieved information and the contexts of use of retrieved information.

**References**

Afolabi, B. (2007). 'La conception et l'adaptation de la structure d'un système d'intelligence économique par l'observation des comportements de l'utilisateur.' Ph.D. Nancy 2 University. France.

Chevallet, J. Nigy, L. (1996). 'Characteristics of users' needs and activities: A design space for interactive information retrieval system'. Glasgow Workshop on information retrieval and human computer interaction. Glasgow. http://iihm.imag.fr/site/publiAnnee_an_1996.html

David, A. (1999). 'Modélisation de l'utilisateur et recherche coopérative d'information dans les systèmes de recherche d'information multimédia en vue de la personnalisation des réponses'. HDR. Nancy 2 University, France.

David, A. (2005). 'La recherche collaborative d'information dans un contexte d'intelligence économique'. Paper presented at the conference le système d'information de l'entreprise, Algérie – Telecom, Algérie.

Kuhlhau, C. (1996). 'The concept of a zone of intervention for identifying the role of intermediaries in the information search process'. Annul conference proceeding of American Society of Information science, ASIS.

Kuhthau, C. (1997), 'the influence of uncertainty on the information seeking behaviour of a securities analyst'. In: p. Vakkari, R. Savolainen, and B. Dervin, eds, information seeking in context : proceedings of an international conference on research in information needs, seeking and use in different contexts, Finland , 1996. London / Taylor Graham, 1997, 467p.

Kurzke, C. Galle, M. Bathelt, M. (1998) 'Web Assist: A User Profile Specific Information Retrieval Assistant'. Computer Networks 30(1-7): 654-655

Maghrebi, H. (2007). 'Proposition d'une approche de représentation et d'exploitation des documents audiovisuels pour l'extraction de connaissance'. Paper presented at the 6ème colloque international du chapitre français de l'ISKO. France.

Maghrebi, H. David, A. (2006). 'Toward a model for the representation of multimedia information based on users' needs: economic intelligence approach'. Paper presented at the IV International conference on Multimedia and Information and Communication Technologies in Education. Spain.

Moores, CN. (1951). 'Zatocoding applied to mechanical organisation of knowledge'. In American documentation, 2, P.20.

Robert, C. (2007). 'L'annotation pour la recherche d'information dans le contexte d'intelligence économique'. Ph.D. Nancy 2 University. France.

Saracevic, T. (1995). 'Evolution of evolution in information retrieval'. SIGIR 95, Seattle CA.

Saracevic, T. (1996). 'Modelling interaction information retrieval: a review and proposal'. In: proceeding of the American Society of Information Sciences, Vol .33, Pc3-9.

Wilson, Tom. (1999). 'Models in information behaviour research'. Journal of documentation, 55(3), P. 249-269.

Wilson, Tom. (2000) 'recent trend in user studies: action research and qualitative methods. Information research'.

**Figure 1.** Integration of user informational needs in the database

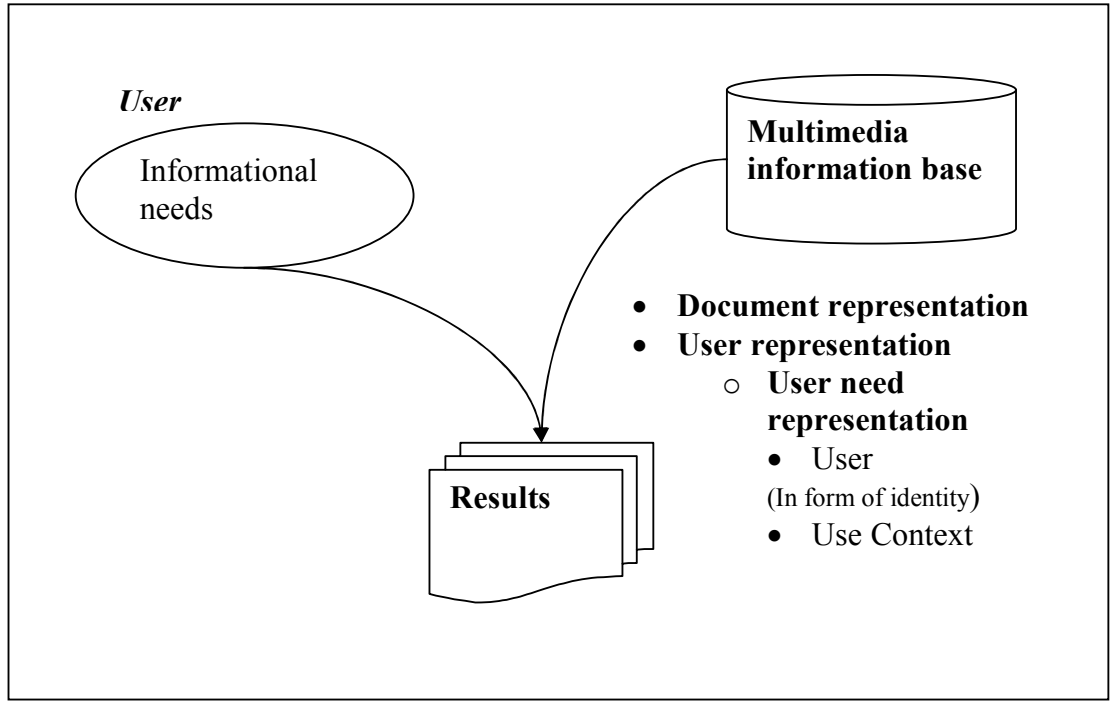

**Figure 2**

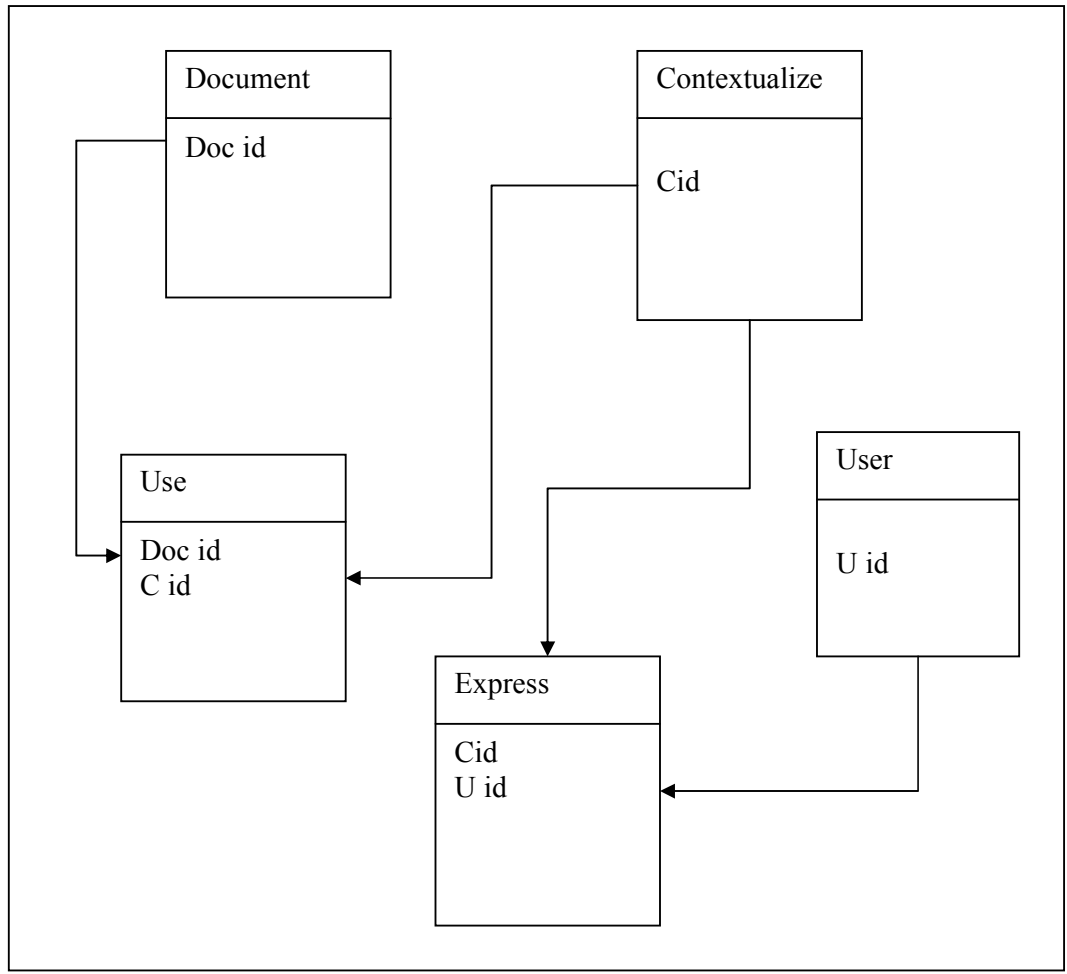